\input{aipcheck}

\documentclass[final]{aipproc}
\layoutstyle{6x9}

\begin{document}

\title{Running coupling constant and masses in QCD, the meson spectrum}
\author{M. Baldicchi and G. M. Prosperi}{address={Dipartimento di
Fisica, Universit\`a di Milano\\
I.N.F.N., sezione di Milano\\
via Celoria 16, I20133 Milano, Italy}}

\begin{abstract}
In line with some previous works, we study
in this paper the meson spectrum
in the framework of a second order quark-antiquark Bethe-Salpeter
formalism which includes confinement. An analytic one loop running
coupling constant $\alpha_{\rm s}(Q)$, as proposed by Shirkov and
Sovlovtsov, is used in the calculations. As for the quark masses,
the case of a purely phenomenological running mass for the light
quarks in terms of the c. m. momentum is further investigated.
Alternatively
a more fundamental expression $m_P(Q)$ is introduced for light and
strange quarks, combining renormalization group and analyticity
requirements with an approximate solution of the Dyson-Schwinger
equation. The use of such running coupling constant and masses turns
out to be essential for a correct reproduction of the the light
pseudoscalar mesons.

\end{abstract}

\maketitle


\section{1. Introduction}

In a series of papers \cite{quadratic,infrared} we have applied a second order
Bethe-Salpeter formalism~\footnote{second order in the sense of the
differential equations}, previously established \cite{bmp}, to the
evaluation of the quark-antiquark spectrum, in the context of QCD.
Taking advantage of a Feynman-Schwinger representation for the quark
propagator in an external field, the kernels of the Bethe-Salpeter and
the Dyson-Schwinger equations were obtained, starting from an appropriate
ansatz on the Wilson loop correlator. Such an ansatz consisted in adding
an area term to the lowest perturbative expression of $\ln W$. 
By a 3D reduction of the original 4D BS equation
a mass operator was obtained 
and applied to the determination of the $q\overline q$ bound
states \cite{simon}. 

In that way, using a fixed strong coupling constant $\alpha_{\rm s}$ and
appropriate values for the other variables, the entire spectrum  was reasonably
well reproduced with, however, the relevant exception of the
light  pseudoscalar mesons ($\pi, K, \eta_s$).
Agreement even for the latter states could be obtained using an
analytic running coupling constant $\alpha_{\rm  s}(Q)$ proposed by
Shirkov and Sovlovtsov, which is modified in the infrared region with
respect to the ordinary purely perturbative expression
\cite{sanda,shirkov}. In conjunction it was also necessary to use a 
phenomenological running constituent mass for the light
quarks $u$ and $d$, written as a polynomial in the center
of the mass quark momentum 
${\bf k}$ \cite{infrared}. 

   In this paper we reconsider and improve the above procedure from
two aspects:

 a) we evaluate the hyperfine $^3S_1 - {^1S}_0$ separation for the light
pseudoscalar mesons to the second rather than to the first order
perturbation theory,

 b) we use running constituent masses for $u,d$ and $s$ quarks,
obtained by an approximate
solution of the appropriate DS equations and
analytic running current masses. 

As a consequence of a) a significant improvement
is obtained in agreement with
the data  already with a phenomenological running
mass for the light quarks. As for case b),
preliminary calculations seem to provide results numerically
similar to the above ones, but more satisfactory from the conceptual
point of view.

The plan of the remaining part of the paper is as follows. In
Sect. 2 we briefly recall the second order BS formalism to
establish notations. In Sect. 3 we discuss the DS equation and the 3D 
reduction of the BS equation. In Sect. 4 we consider the infrared
behavior of the running coupling constant and obtain the
corresponding running masses. In sect. 5 and 6 we report our
results and draw some conclusions.


\section{2. Second order Bethe-Salpeter formalism}

    In the QCD framework a {\it second order} four point quark-antiquark
function and a full quark propagator can be defined as
\begin{equation}
H^{(4)}(x_1,x_2;y_1,y_2)= -{1\over 3} {\rm Tr _{color}}
\langle \Delta_1 (x_1,y_1;A)
{\Delta}_2(y_2,x_2;A)\rangle 
\label{eq:so4point}
\end{equation}
\noindent
and
\begin{equation}
H^{(2)}(x-y) = {i \over \sqrt{3}}{\rm Tr_{color}}
\langle \Delta(x,y:A)\rangle \,,
\label{eq:so2point}
\end{equation}
where
\begin{equation}
\langle f[A] \rangle = \int  DA\, M_F [A]\, e^{iS_G[A]} f[A]  \,,
\label{eq:expt}
\end{equation}
\noindent
$
M_F[A] = {\rm Det} \, \Pi_{j=1}^2 [1 + g\gamma^\mu A_\mu
( i\gamma_j^\nu \partial_{j\nu} - m_j)^{-1}]
$
and $\Delta (x,y;A)$ is the {\it second order} quark
propagator in an external gauge field. 

The quantity $ \Delta $ is defined by
the second order differential equation
\begin{equation}
(D_\mu D^\mu +m^2 -{1\over 2} g \, \sigma^{\mu \nu} F_{\mu \nu})
\Delta (x,y;A) = -\delta^4(x-y) \, ,
\label{eq:soprop}
\end{equation}
($\sigma^{\mu \nu} = {i\over 2} [\gamma^\mu, \gamma^\nu]$ and 
$D_\mu=\partial_\mu + ig A_\mu$) and it is related to
the corresponding first order propagator by 
$
S(x,y;A) = (i \gamma^\nu D_\nu + m) \Delta (x,y;A) \,.
$
%

   The advantage of considering second order quantities is that the spin
terms are more clearly separated and it is
possible to write for $\Delta$ a generalized Feynman-Schwinger
representation, {\it i. e.} to solve eq. (\ref{eq:soprop})
in terms of a quark path integral \cite{bmp,quadratic}. 
Using the latter in (\ref{eq:so4point}) or (\ref{eq:so2point}) a similar
representation can be obtained for $H^{(4)}$ and $H^{(2)}$.

The interesting aspect of this representation is that the gauge field
appears only through a Wilson line correlator $W$.
  In the limit $x_2 \to x_1$, $y_2 \to y_1$ or $y \to x$ the Wilson lines
close in a single Wilson loop $\Gamma$
%
%
and if $\Gamma$ stays on a plane,
$i\ln W$ can be written 
in a first approximation
as the sum of its lowest perturbative
expression and an area term
\begin{eqnarray}
&& i\ln W = {4\over 3} g^2 \oint dz^\mu \oint dz^{\nu \prime}
D_{\mu \nu}(z-z^\prime) +
\label{eq:wilson} \\
&& \sigma \oint dz^0 \oint dz^{0 \prime} \delta (z^0-z^{0\prime})
|{\bf z} - {\bf z}^\prime| \int_0^1 d\lambda
 \Big \{ 1 -  [\lambda {d{\bf z}_\perp \over dz^0}
 + (1-\lambda) {d{\bf z}_\perp ^\prime \over dz^{0 \prime}} ]^2 
\Big \}^{1\over 2} \, . \nonumber
\end{eqnarray}
The area term here is written as the algebraic sum of successive
equal time strips and $ d{\bf z}_\perp = d{\bf z} - 
(d{\bf z}\cdot {\bf r}){\bf r}/r^2 $ denotes the transversal component of
$ d{\bf z} $. The basic assumption now
is that in the center of mass frame 
(\ref{eq:wilson}) remains a good approximation even
in the general case, {\it i. e.}
for non flat curves and  when $x_2 \ne x_1$, 
$y_2 \ne y_1$ or $y \ne x$.

Then, by appropriate manipulations on the resulting expressions,
an inhomogeneous
Bethe-Salpeter equation for the 4-point function
$H^{(4)}(x_1,x_2;y_1,y_2)$ and a Dyson-Schwinger equation for
$H^{(2)}(x-y)$ can be derived in a kind of generalized ladder and rainbow
approximation. This should appear plausible, even from the point of
view of graph resummation, for the analogy between the perturbative
and the confinement terms in (\ref{eq:wilson}).
We may refer to such terms
as a {\it gluon exchange} and a {\it string connection}.

   In momentum representation, the corresponding homogeneous
BS-equation becomes
\begin{eqnarray}
\Phi_P (k) &=& -i \int {d^4u \over (2 \pi)^4} \;
   \hat I_{ab} \left( k-u; \, {1 \over 2}P
   +{k+u \over 2}, \,
   {1 \over 2}P-{k+u \over 2} \right) \nonumber \\
    & & \qquad \qquad
   \hat H_{1}^{(2)}   \left({1 \over 2} P  + k \right)
      \sigma^a  \, \Phi_P (u) \, \sigma^b \,
   \hat H_{2}^{(2)} \left(-{1 \over 2} P + k \right) \, ,
\label{eq:bshom}
\end{eqnarray}
\noindent
where we have set $\sigma^0=1$; $a, \, b = 0, \, \mu\nu$;
the center of mass frame has to be understood, $P=(m_B, {\bf 0})$;
$\Phi_P (k)$ denotes an appropriate {\it second order} wave function
\footnote{In terms of the second order field $\phi (x) = (i\gamma^\mu
D_\mu + m)^{-1}\psi(x)$ this wave function is defined by
\[
  \langle 0|\phi({\xi \over 2}) \bar\psi(-{\xi \over 2}) |P\rangle  =
  {1\over (2\pi)^2} \Phi_P (k) e^{-ik\xi} \,.
\] }.

  Similarly, in terms of the irreducible self-energy, defined by
$\hat H^{(2)}(k) ={i\over k^2-m^2} + {i\over k^2-m^2} \, i \,
\Gamma (k) \, \hat H^{(2)}(k) \,$,
the DS-equation can be written
\begin{equation}
\hat \Gamma(k) =  \int {d^4 l \over (2 \pi)^4}  \,
\hat I_{ab} \Big ( k-l;{k+l \over 2},{k+l \over 2} \Big )
\sigma^a \hat H^{(2)}(l) \, \sigma^b \ .
\label{eq:dshom}
\end{equation}

   The kernels in
(\ref{eq:bshom}) and (\ref{eq:dshom})
are the same in the two equations, consistently with the
requirement of chiral symmetry limit \cite{chiral}, and are given by
\begin{eqnarray}
& & \hat I_{0;0} (Q; p, p^\prime)  =
   16 \pi {4 \over 3} \alpha_{\rm s} p^\alpha p^{\prime \beta}
  \hat D_{\alpha \beta} (Q)  + \nonumber \\
& &  \quad + 4 \sigma  \int \! d^3 {\bf \zeta} e^{-i{\bf Q}
   \cdot {\bf \zeta}}
    \vert {\bf \zeta} \vert \epsilon (p_0) \epsilon ( p_0^\prime )
   \int_0^1 \! d \lambda \{ p_0^2 p_0^{\prime 2} -
   [\lambda p_0^\prime {\bf p}_{\rm T} +
   (1-\lambda) p_0 {\bf p}_{\rm T}^\prime ]^2 \} ^{1 \over 2} \nonumber \\
& & \hat I_{\mu \nu ; 0}(Q;p,p^\prime) = 4\pi i {4 \over 3} \alpha_{\rm s}
   (\delta_\mu^\alpha Q_\nu - \delta_\nu^\alpha Q_\mu) p_\beta^\prime
   \hat D_{\alpha \beta}(Q)  - \nonumber \\
& & \qquad \qquad \qquad  - \sigma  \int d^3 {\bf \zeta} \, e^{-i {\bf Q}
\cdot {\bf \zeta}} \epsilon (p_0)
   {\zeta_\mu p_\nu -\zeta_\nu p_\mu \over
   \vert {\bf \zeta} \vert \sqrt{p_0^2-{\bf p}_{\rm T}^2}}
   p_0^\prime  \nonumber \\
& & \hat I_{0; \rho \sigma}(Q;p,p^\prime) =
   -4 \pi i{4 \over 3} \alpha_{\rm s}
   p^\alpha (\delta_\rho^\beta Q_\sigma - \delta_\sigma^\beta Q_\rho)
   \hat D_{\alpha \beta}(Q) + \nonumber \\
& & \qquad \qquad  \qquad  + \sigma  \int d^3 {\bf \zeta} \, e^{-i{\bf Q}
  \cdot {\bf \zeta}} p_0
  {\zeta_\rho p_\sigma^\prime - \zeta_\sigma p_\rho^\prime \over
  \vert {\bf \zeta} \vert \sqrt{p_0^{\prime 2}
   -{\bf p}_{\rm T}^{\prime 2}} }
  \epsilon (p_0^\prime)  \nonumber \\
& & \hat I_{\mu \nu ; \rho \sigma}(Q;p,p^\prime) =
   \pi {4\over 3} \alpha_{\rm s}
  (\delta_\mu^\alpha Q_\nu - \delta_\nu^\alpha Q_\mu)
  (\delta_\rho^\alpha Q_\sigma - \delta_\sigma^\alpha Q_\rho)
  \hat D_{\alpha \beta}(Q) \, ,
\label{eq:imom}
\end{eqnarray}
\noindent
where in the second and in the third equation $\zeta_0 = 0$ has to be
understood. Notice that, due to the privileged role given to the c. m. frame,
the terms proportional to $\sigma$ in (\ref{eq:imom}) are not formally
covariant.


\section{3. DS equation and mass operator}

Concerning eq. (\ref{eq:dshom}), let us observe that the unity matrix,
$\sigma^{\mu\nu}$ and  $\gamma^5$ form a subalgebra of the Dirac
algebra. Consequently $\Gamma(k)$ can be assumed to depend only on this set
of matrices and, since it must be a three dimensional scalar, only on terms 
like $k_j \sigma^{0j}$. In fact, it can be checked that $\Gamma(k)$ can
be consistently assumed to be completely spin independent and 
eq. (\ref{eq:dshom}) can be written in the form 
\begin{equation}
\Gamma (k) = i \int {d^4l \over (2 \pi)^4} \,
  {R(k,l) \over l^2-m^2+ \Gamma(l)} ,
\label{eq:dssc}
\end{equation}
\noindent
with
\begin{eqnarray}
R(k,l)&=& 4\pi \, {4 \over 3} \, \alpha_{\rm s}\left
 [4 {p^2 l^2 -(pl)^2 \over
      (k-l)^2} + {3 \over 4}\right] + \nonumber \\
  & & + \, \sigma \int d^3 {\bf r} \, e^{-i({\bf k}-{\bf l})
      \cdot{\bf r}} \, r \, (k_0 + l_0)^2 \sqrt{1-{({\bf k_\perp}+{\bf
      l_\perp})^2 \over (k_0 + l_0)^2}} \,,
\label{eq:dskernel}
\end{eqnarray}
\noindent
${\bf k_\perp}$ and ${\bf l_\perp}$ denoting as above the transversal
part of ${\bf k}$ and ${\bf l}$.

Notice that, once (\ref{eq:dskernel}) is solved, the pole or
constituent mass $m_P ,$ to be used in bound states problems, is given
by the equation
\begin{equation}
m_P^2 - m^2 + \Gamma (m_P^2) = 0 \, .
\label{eq:polemass}
\end{equation}

We can try to solve eq. (\ref{eq:dssc}) iteratively and we have at
the first step
\begin{equation}
\Gamma (k) = i \int {d^4l \over (2 \pi)^4} \, {R(k,l) \over l^2-m^2}
\, .
\label{eq:iter}
\end{equation}

In a preliminary calculation we omit altogether
the perturbative contribution to $R(k,l)$ (notice the overplacing
of curves {\it b} and {\it c} in Fig. 1) and neglect the term in
$({\bf k}_\perp + {\bf l}_\perp)^2$ in the string part.
Strictly, the second approximation
is justified only for $S$ bound states (classically 
${k_\perp}r$ is the angular momentum of the bound state)
but it is necessary in order to make the integral analytically
calculable.

Then introducing a cut off $ \mu $, we obtain
\begin{equation}
  \Gamma (k) = {\sigma \over \pi}[k_0^2 A(m,|{\bf k}|)
  - B_{\mu} - B(m,|{\bf k}|) 
\,, \label{eq:Gexpl}
\end{equation}
\noindent 
where $ B_{\mu} = 2\ln { {\mu} \over m} - 1$,  
\begin{equation}
 A(m,|{\bf k}|) = {1 \over {\bf k}^2 + m^2}\left [1 + {m^2 \over 
2|{\bf k}| \sqrt{{\bf k}^2 + m^2}} \ln {\sqrt{{\bf k}^2 + m^2} + 
|{\bf k}| \over \sqrt{{\bf k}^2 + m^2} - |{\bf k}|} \right ]
\label{eq:coeff}
\end{equation}
\noindent
and $B(m,|{\bf k}|)$ is a more complicated expression that we do not
report explicitly here for lack of space. The resulting pole mass is
\begin{equation}
\overline m_P^2 (m,|{\bf k}|) = {m^2 + {\sigma \over \pi}[B_{\mu} +
B(m,|{\bf k}|)-{\bf k}^2 A(m,|{\bf k}|)]\over 1+ {\sigma \over \pi} 
A(m,|{\bf k}|)}
\,, \label{eq:cnstmass}
\end{equation}

The above expression depends on the current mass $m$ and 
on the quark c. m. momentum $|{\bf k}|$, 
(see Fig. 1{\it a}). Notice that such dependence on  $|{\bf k}|$ is
clearly an artifact of the schematic  way we have introduced confinement
in eq. (\ref{eq:wilson}) and that the curve is rather flat in the region of
interest. Correspondingly it seems reasonable to chose as true mass
$m_P(m)$
the value of $\overline m_P (m,|{\bf k}|)$ at its stationary point
in $|{\bf k}|$.

Then, in a neighbor of its singularity $k^2=m_P^2$, the full propagator
can be written as $\hat H^{(2)}(k) = {i \, Z \over k^2 - m_P^2}$, where
the residuum $Z$ differs from 1 only for terms proportional to
$\alpha_s$ or $\sigma$. Consistently in (\ref{eq:bshom}) we
can simply take $Z=1$ and are left with the free propagator with a 
constituent mass. If, in addition, we replace $\hat{I}_{ab}$ with its
so called instantaneous approximation $ \hat{I}_{ab}^{\rm inst}
({\bf k}, {\bf u})$, we can explicitly perform the integration in
$u_0$ and arrive at a three dimensional reduced equation.     
 
  Such a reduced equation takes the form of the eigenvalue equation
for a squared mass operator \cite{bmp},
$
    M^2 = M_0^2 + U \,,
$
with  $ M_0 = w_1 + w_2 $, $w_{1,2}=\sqrt{m_{1,2}^2 + {\bf k}^2}$  and
\begin{equation}
   \langle {\bf k} \vert U \vert {\bf k}^\prime \rangle =
        {1\over (2 \pi)^3 }
        \sqrt{ w_1 + w_2 \over 2  w_1  w_2} \; \hat I_{ab}^{\rm \; inst}
        ({\bf k} , {\bf k}^\prime) \; \sqrt{ w_1^\prime + w_2^\prime \over 2
         w_1^\prime w_2^\prime}\; \sigma_1^a \sigma_2^b \,
\label{eq:quadrrel}
\end{equation}
(for an explicit expression we refer to \cite{infrared,quadratic}).
The quadratic form of the above equation obviously derives from the second
order formalism we have used.

  Alternatively, in more usual terms, one can look for the eigenvalue of the
mass operator or center of mass Hamiltonian
$          H_{\rm CM} \equiv M = M_0 + V  $
with $V$ defined by $M_0V+VM_0+V^2=U$. Neglecting the term $V^2$
the linear form potential $V$ can be
obtained from $U$ by
the kinematic replacement
$
\sqrt{ (w_1+w_2) (w_1^\prime +w_2^\prime)\over w_1w_2w_1^\prime w_2^\prime}
\to {1\over 2\sqrt{w_1 w_2 w_1^\prime w_2^\prime}}
$.
The resulting
expression is particularly useful for a comparison with models based
on potential. In particular, in the static limit $V$ reduces to the Cornell
potential
\begin{equation}
V_{\rm stat} = - {4 \over 3} {\alpha_{\rm s} \over r } + \sigma r \, ;
    \label{eq:static}
\end{equation}
in the semirelativistic limit (up to ${1 \over m^2}$ terms after an
appropriate Foldy-Wouthuysen transformation) it equals the
potential discussed in ref. \cite{barch}, if 
full relativistic kinematics is kept,
but the spin dependent terms are neglected, it becomes
identical to the potential of the relativistic flux tube model \cite{bmp}.


\section{4. Running coupling constant and masses}

As we said, diagonalizing $M^2$ or $H_{\rm CM}$  with fixed coupling
constant and quark masses, a general good fit of the data was obtained.
Actually a serious problem was represented by the masses of the light
pseudo scalar mesons that turned out too large.
The results obtained in ref. \cite{infrared} suggest,
however, that the situation can be greatly improved using an
appropriate running coupling constant and running quark masses.

At one loop, the running coupling constant is usually written
\begin{equation}
  \alpha_{\rm s} ( Q ) = \frac{ 4 \pi }{ \beta_{0}
  \ln{ ( Q^{2} / \Lambda^{2} ) } } \, ,
\label{eq:runcst}
\end{equation}
with $\beta_0=11-{2 \over 3} N_{\rm f}$ and $ N_{\rm f} $ the number of
`active' quarks. However,
the singularity occurring in such expression is an
artifact of perturbation theory
and it contradicts general analyticity properties, therefore the
expression must be somewhat modified in the infrared region
\cite{ginzburg}.
Notice that this is particularly important for the quark-antiquark
bound state problem, where the variable $ Q^{2} $ is usually
identified with the squared momentum transfer
$ {\bf Q}^{2} = ( {\bf k} - {\bf k}^{\prime} )^{2} \! $,
which ranges typically from
$ \; ( 0.1 \, {\rm GeV} )^{2} $ to
$ ( 1 \, {\rm GeV} )^{2} $
for different quark masses and states.

The most naive modification of eq. (\ref{eq:runcst}) would
consist in freezing $\alpha_{\rm s} (Q^{2})$
to a certain maximum value $\bar{\alpha}_{\rm s}$ as
$Q^2$ decreases
and in treating this value as a phenomenological parameter
(truncation prescription).
However, various more sensible proposals have been made
on different bases
\cite{sanda,shirkov}.

In particular Shirkov and Solovtsov \cite{shirkov} suggest
to replace (\ref{eq:runcst}) with
\begin{equation}
  \alpha_{\rm s} ( Q ) = \frac{ 4 \pi }{
   \beta_{0} } \left(
  \frac{1}{ \ln{ ( Q^{2} / \Lambda^{2} ) } } +
  \frac{ \Lambda^{2} }{ \Lambda^{2} - Q^{2} } \right).
\label{eq:runshk}
\end{equation}
This remains regular for $ Q^{2} = \Lambda^{2} $ and
has a finite $ \Lambda $ independent limit
$ \alpha_{\rm s}(0) = 4 \pi / \beta_{0} $
for $ Q^{2} \rightarrow 0 $. Eq. (\ref{eq:runshk}) is
obtained assuming a dispersion relation for
$ \alpha_{\rm s} ( Q ) $
with a cut for
$-\infty<Q^2<0$ and applying
(\ref{eq:runcst}) to
the evaluation of the
spectral function.


The running mass expression corresponding to (\ref{eq:runshk}) can
be written in the form
\begin{equation}
  m ( Q ) = \hat m \, \left( {Q^2/\Lambda^2 - 1 \over Q^2/\Lambda^2 
\ln (Q^2/\Lambda^2) } \right)^{\gamma_0/2\beta_0} \, ,
\label{eq:runmass}
\end{equation}
\noindent
where in the $\overline {MS}$ scheme $\gamma_0=8$.
Eq. (\ref{eq:runmass})
 is obtained integrating the one loop renormalization
group equation 
\begin{equation}
  {Q \over m ( Q )} {dm( Q ) \over dQ} = - \gamma_0 
{\alpha_{\rm s} ( Q ) \over 4 \pi} \, ,
\label{eq:rngr}
\end{equation}
\noindent
where (\ref{eq:runshk}) has been used
and $\hat m$ denotes an
integration constant. Notice that
$ m ( Q )$ is singular for $ Q \rightarrow 0 $,
contrary to $\alpha_{\rm s}(Q)$.

Finally, if we replace the running mass (\ref{eq:runmass}) in 
(\ref{eq:cnstmass})
after maximizing
we obtain a running constituent mass $ m_P (Q)$
of the type reported in Fig. 1 {\it d} that can be used together with the
running coupling constant (\ref{eq:runshk}) in the expression of the
operator $M^2$ (see Sec. 3).
\footnote{At first sight it could seem strange that we should
talk of a $Q$
dependence for a quantity like the constituent mass that should have a
definite physical value. The point is that we are using
$m_P(Q)$
in the context of certain approximations and it is
the accuracy of such approximations that depend on the scale
$Q$.}


\section{5. Calculations and results}

The calculations we report in this paper follow a similar line to those
of Ref. \cite{quadratic}. The general strategy for solving the eigenvalue
equation for $ M^2 $ and the numerical
treatment are basically the same. 

We neglect spin-orbit terms, but include the hyperfine terms
in $U$ (see eq. (\ref{eq:quadrrel})).
We solve first the eigenvalue equation for 
$M_{\rm stat}=M_0+V_{\rm stat}$ (see Eq. (\ref{eq:static})) by the
Rayleigh-Ritz method with an harmonic oscillator basis and then treat 
$M^2 - M_{\rm stat}^2$ as a perturbation (up
to the first order this is
obviously equivalent to taking $m_{\rm B}^2 = \langle M^2 \rangle $).

In the above general framework, in Fig. 2 we graphically report and compare
with the data \cite{data} three different type of results,
corresponding to different choices for the strong coupling
constant $\alpha_{\rm s}$, the string tension $\sigma$ and the
constituent masses.

Diamonds correspond to results already reported in \cite{quadratic}.
A running coupling constant $\alpha_{\rm s}(Q)$ was
assumed equal to the one loop perturbative expression
(\ref{eq:runcst})
frozen at the maximum value $\overline \alpha_{\rm s} = 0.35$,
with $N_f=4$ and $\Lambda = 200$ MeV. $Q$ was identified
with $|{\bf k}- {\bf k}^\prime|$  and $\sigma $ was set equal to
$0.2 \ {\rm GeV}^2$. Fixed  masses $m_u=m_d= 10$ MeV, $m_s = 200$ MeV,
$m_c=1.394$ GeV, $ \; m_b=4.763$ GeV were adopted.
The results do not differ essentially from
the fixed coupling constant case;
the spectrum is reasonably well reproduced 
on the whole with the
exception of the light pseudoscalar mesons
$\pi$, $\eta_s$ and $K$ (the $\eta_s$ mass is
derived from the masses of $\eta$ and $\eta^\prime$
with the usual assumptions).

\begin{figure}
  \begin{picture}(435,150)
    \put(20,15){\includegraphics[height=.22\textheight,
                      width=.3\textheight]{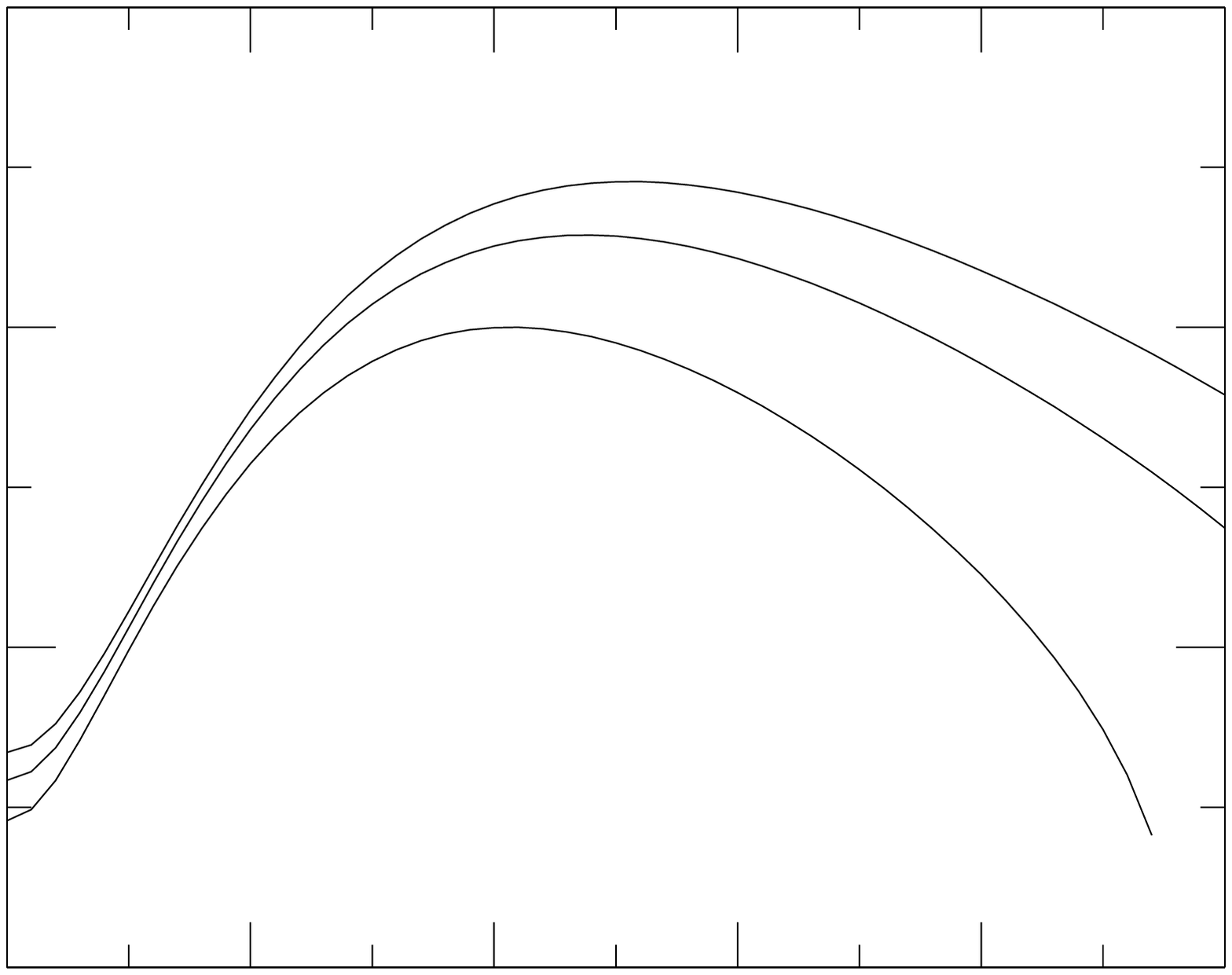}}
    \put(240,15){\includegraphics[height=.22\textheight,
                      width=.3\textheight]{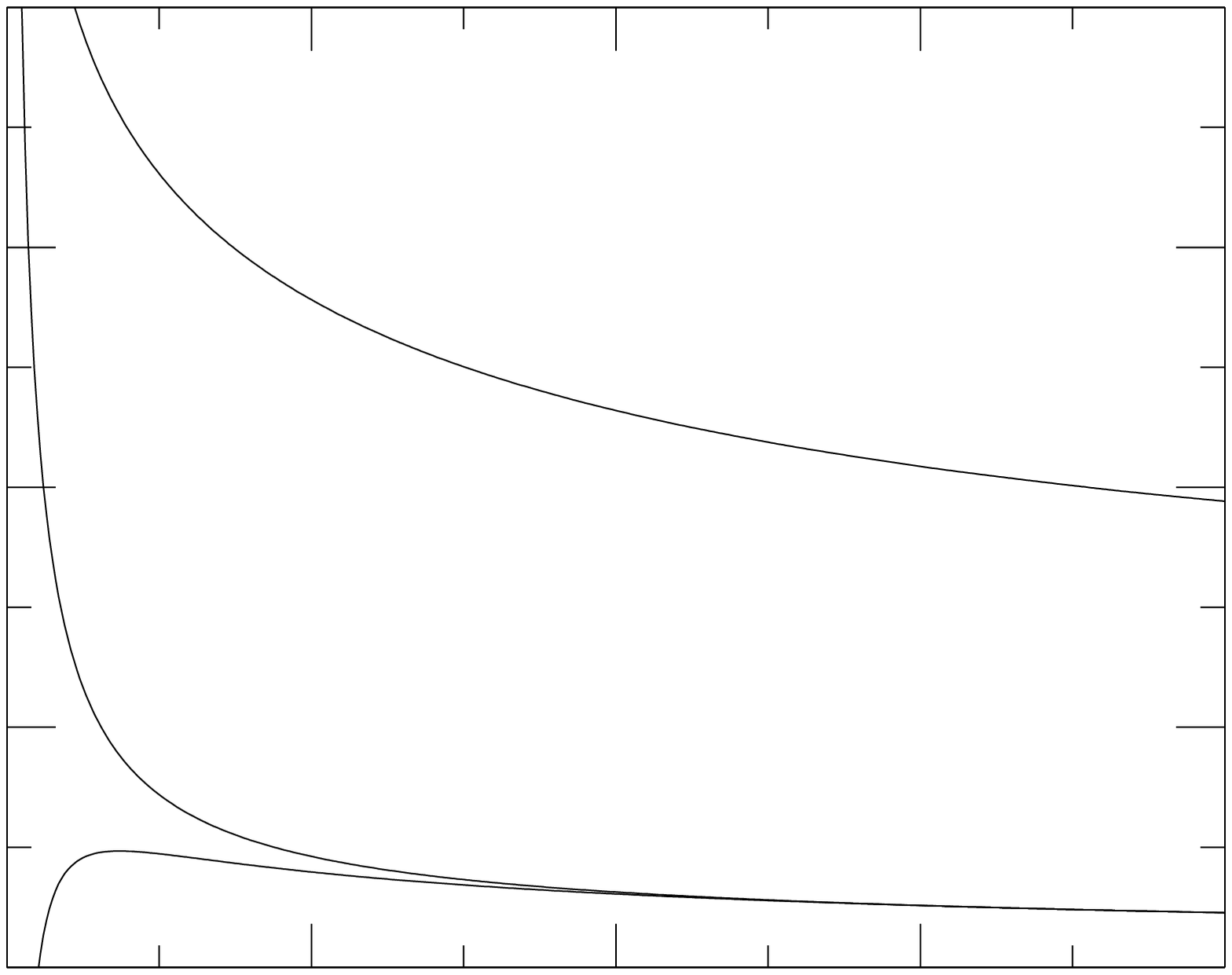}}
      \put(177,115){ {\it a''} }
      \put(177,75){ {\it a'} }
      \put(177,35){ {\it a} }
      \put(10,165){ $ {}_{ ( {\rm GeV} ) } $ }
      \put(2,151){ $ {}_{ 0.3 } $ }
      \put(2,107){ $ {}_{ 0.2 } $ }
      \put(2,62){ $ {}_{ 0.1 } $ }
      \put(190,8){ $ {}_{ ( {\rm GeV} ) } $ }
      \put(177,8){ $ {}_{ | {\bf k} | } $ }
      \put(10,13){ $ {}_{ 0 } $ }
      \put(123,8){ $ {}_{ 0.3 } $ }
      \put(86,8){ $ {}_{ 0.2 } $ }
      \put(49,8){ $ {}_{ 0.1 } $ }
      \put(290,113){ {\it d} }
      \put(255,75){ {\it c} }
      \put(250,20){ {\it b} }
      \put(230,165){ $ {}_{ ( {\rm GeV} ) } $ }
      \put(222,151){ $ {}_{ 0.4 } $ }
      \put(222,118){ $ {}_{ 0.3 } $ }
      \put(222,85){ $ {}_{ 0.2 } $ }
      \put(222,51){ $ {}_{ 0.1 } $ }
      \put(410,8){ $ {}_{ ( {\rm GeV} ) } $ }
      \put(400,8){ $ {}_{ {\it Q} } $ }
      \put(230,13){ $ {}_{ 0 } $ }
      \put(371,8){ $ {}_{ 0.3 } $ }
      \put(325,8){ $ {}_{ 0.2 } $ }
      \put(279,8){ $ {}_{ 0.1 } $ }
  \end{picture}
  \caption{{\it a}), {\it a'}), {\it a''})
$ \; {\bar m}_P = {\bar m}_P (m, |{\bf k}|) $
for increasing value of $m$ (16);
{\it d}) running constituent mass $ m_{\rm P} (Q) $;
{\it c}) analytic running current mass (21);
{\it b}) perturbative running pole masses 
corresponding to {\it c}).}
\end{figure}
\begin{figure}[htbp!]
 \begin{picture}(420,600)
 \put(20,450){\includegraphics[height=.22\textheight,
                      width=.3\textheight]{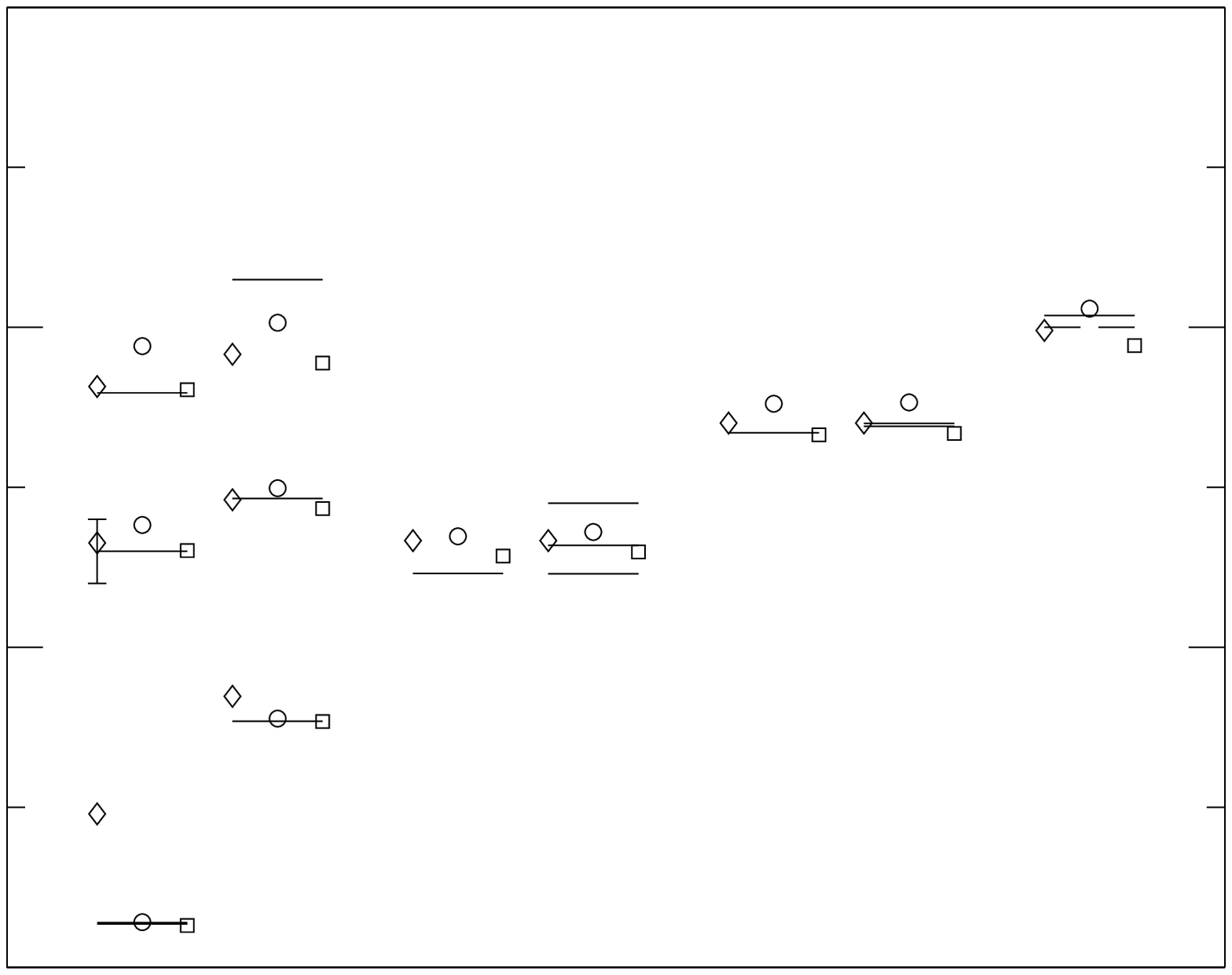}}
 \put(162,118){\includegraphics[height=.82\textheight,
                      width=.6\textheight]{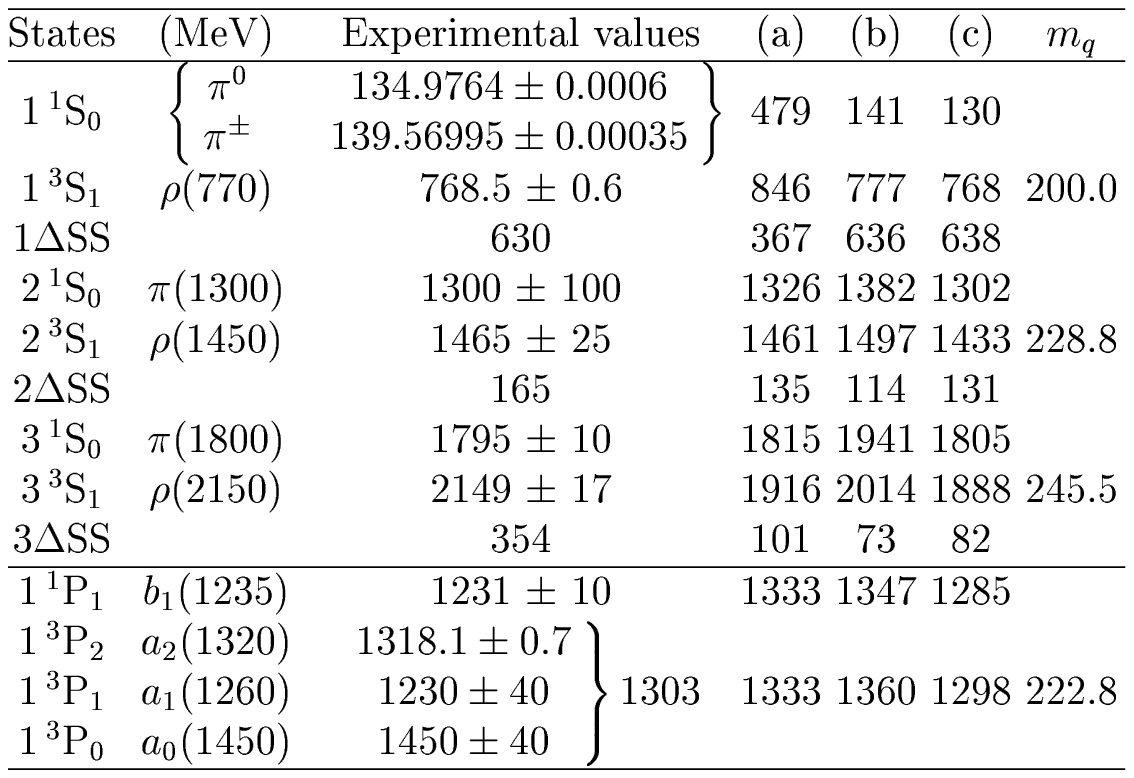}}
 \put(20,300){\includegraphics[height=.22\textheight,
                      width=.3\textheight]{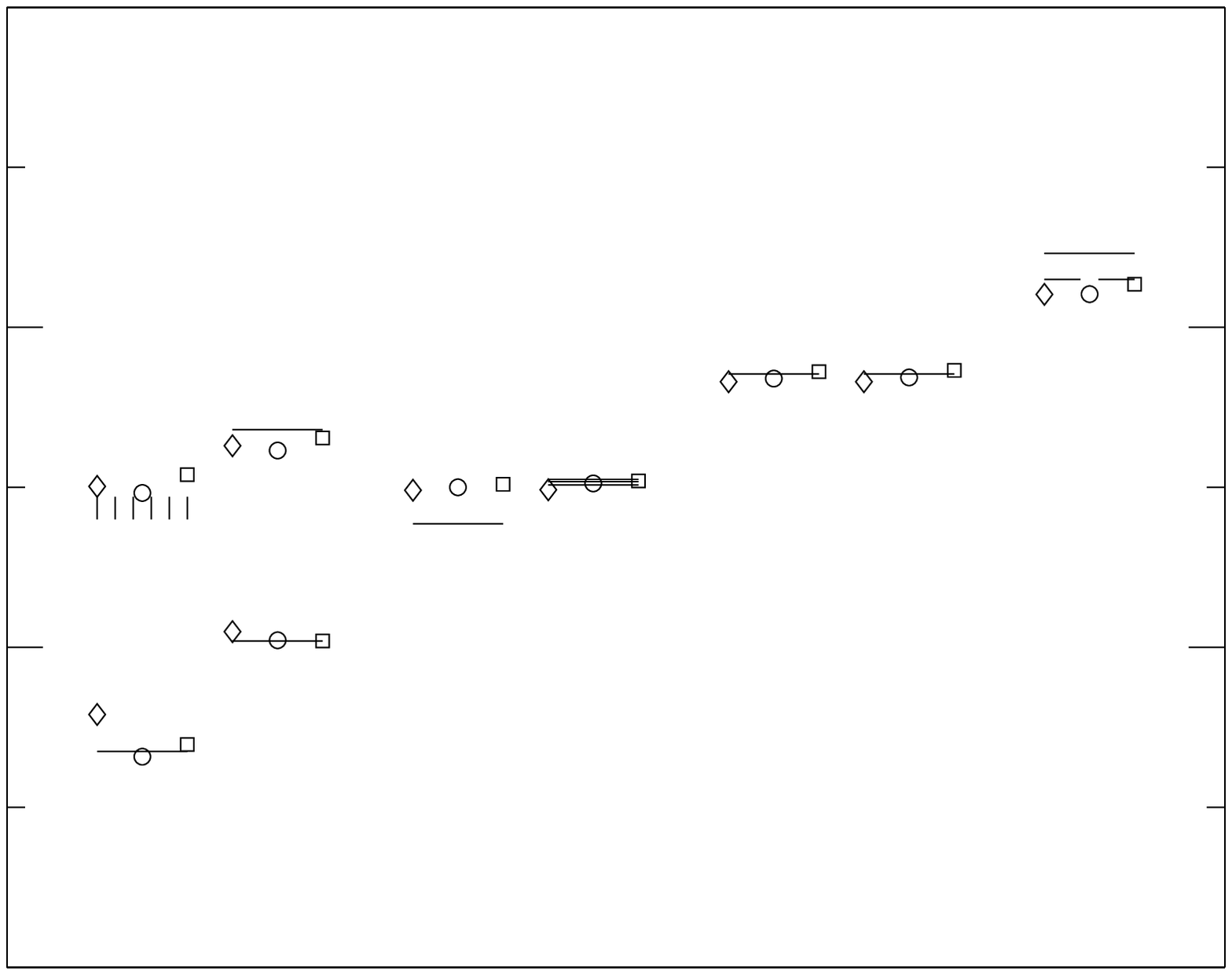}}
 \put(235,300){\includegraphics[height=.22\textheight,
                      width=.3\textheight]{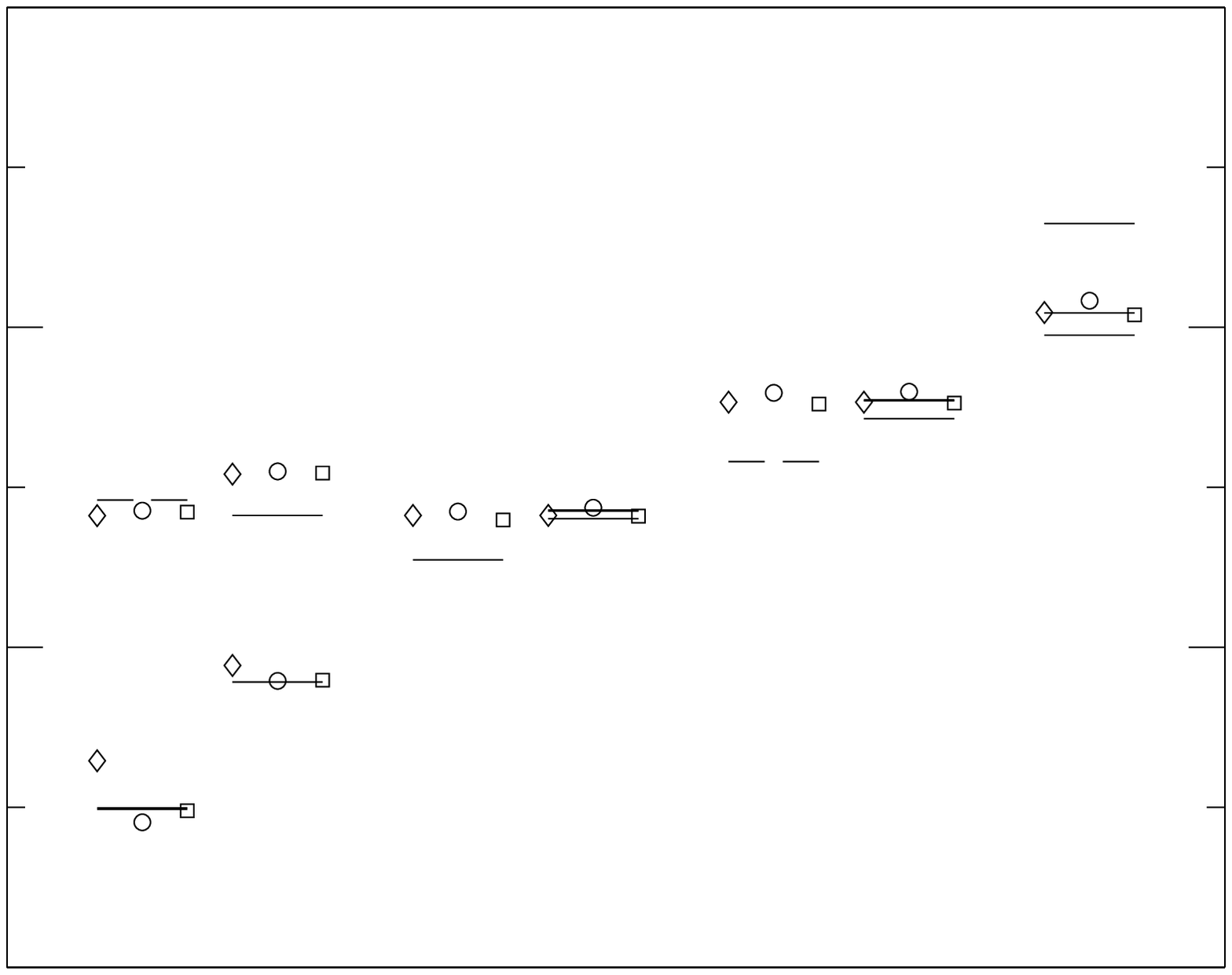}}
 \put(20,150){\includegraphics[height=.22\textheight,
                      width=.3\textheight]{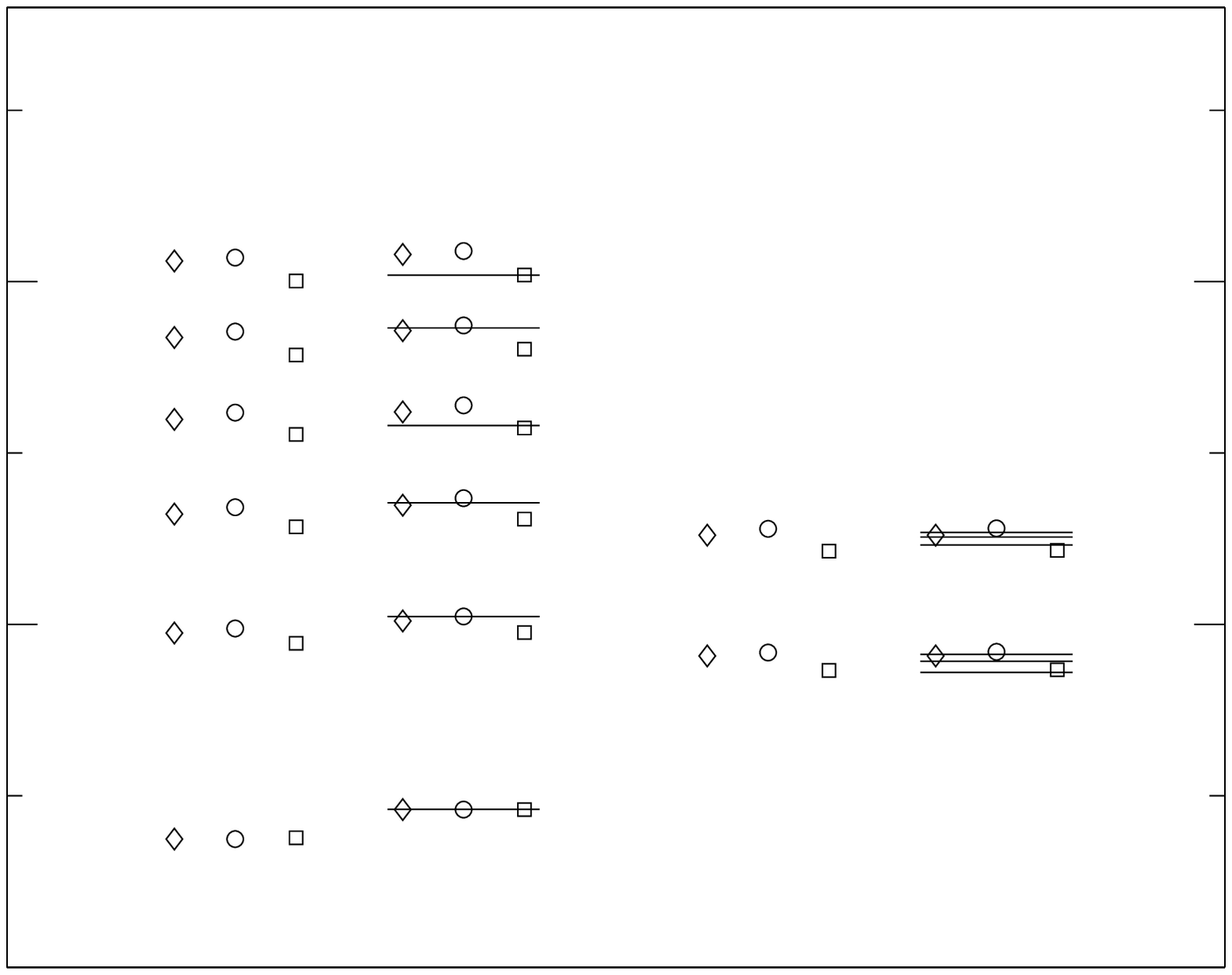}}
 \put(235,150){\includegraphics[height=.22\textheight,
                      width=.3\textheight]{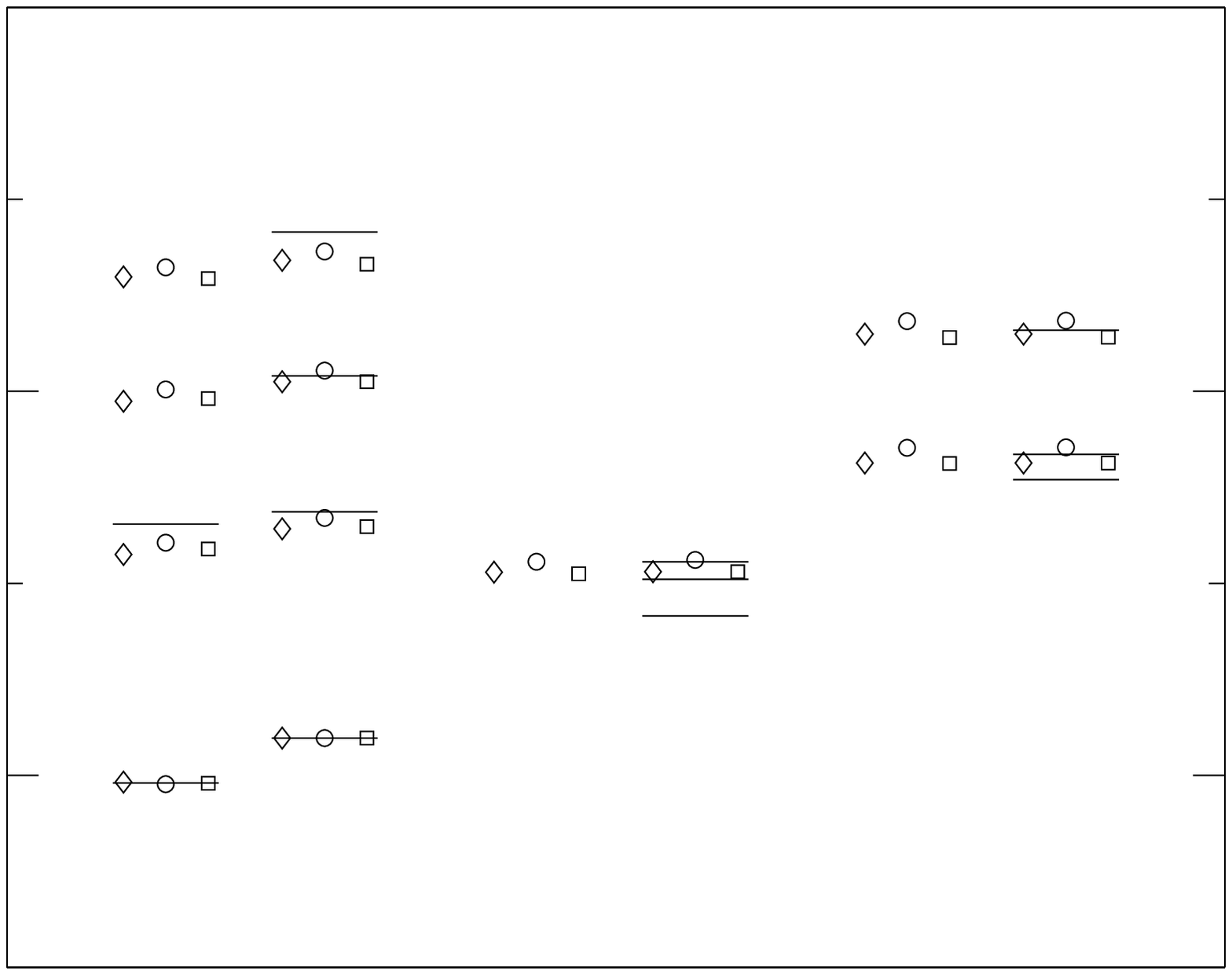}}
 \put(20,0){\includegraphics[height=.22\textheight,
                      width=.3\textheight]{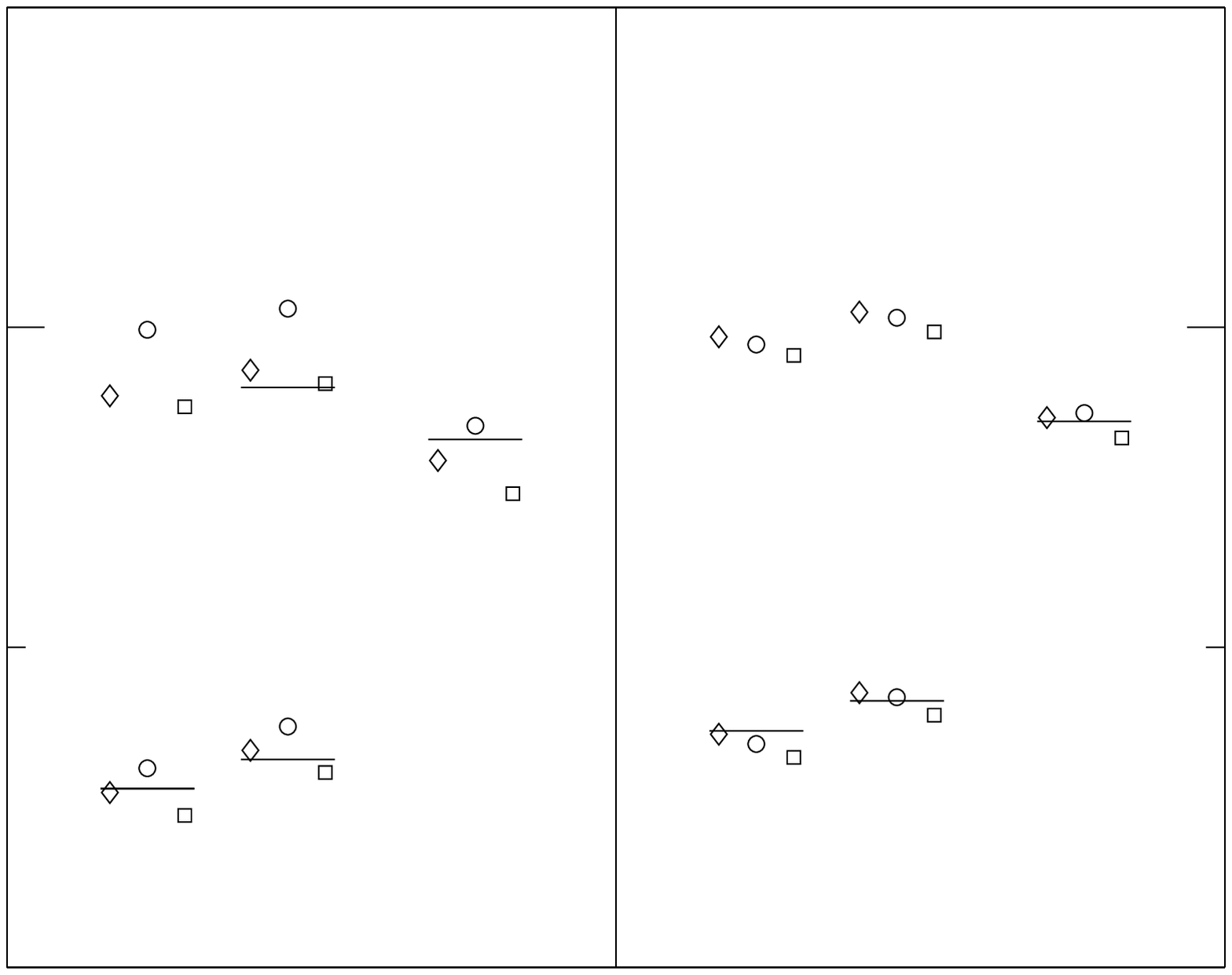}}
 \put(235,0){\includegraphics[height=.22\textheight,
                      width=.3\textheight]{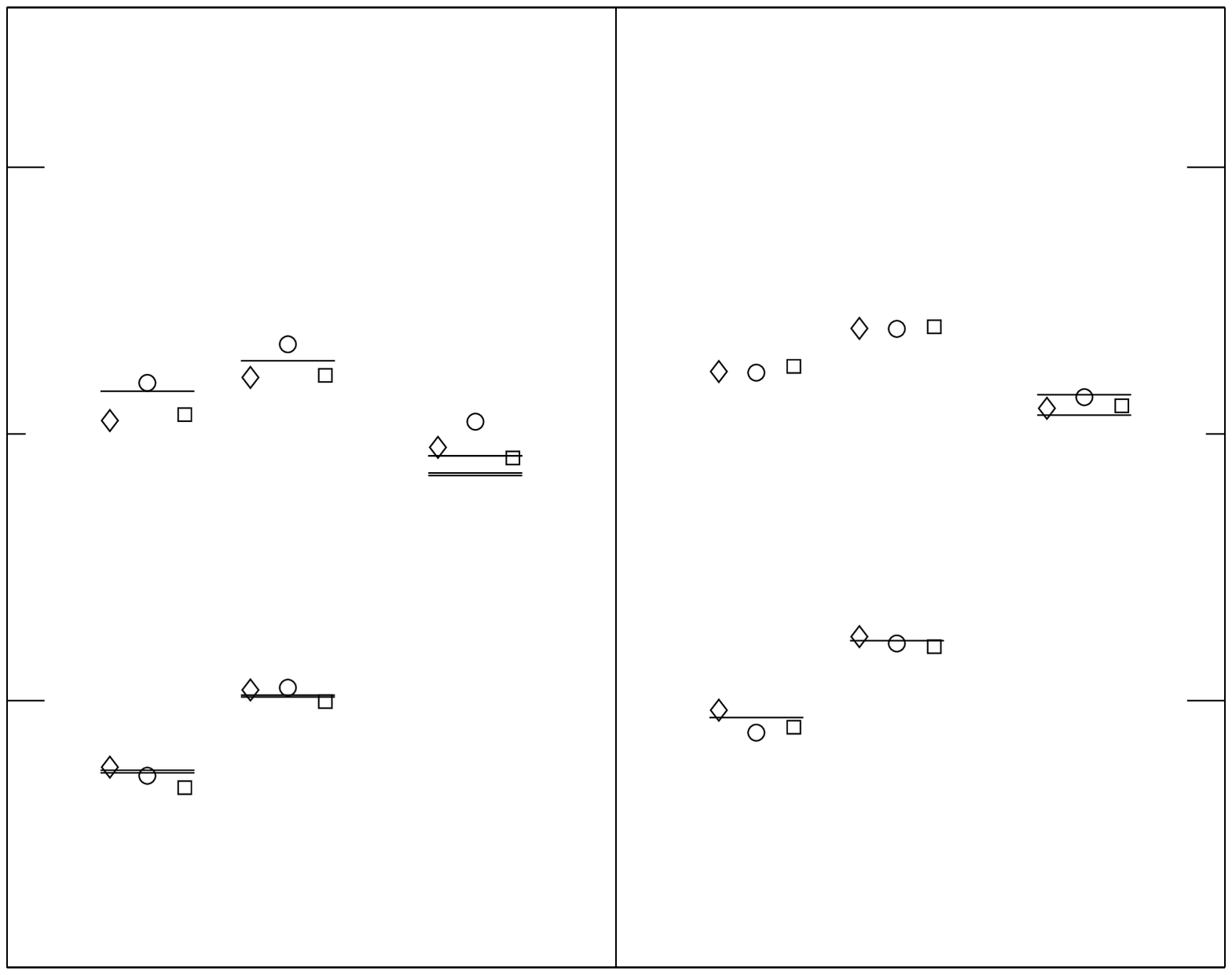}}
      \put(5,600){ $ {}_{ ( {\rm MeV} ) } $ }
      \put(0,587){$ {}_{3000} $}
      \put(0,542){$ {}_{2000} $}
      \put(0,497){$ {}_{1000} $}
      \put(13,452){$ {}_{0} $}
      \put(160,475){ $ q \bar{q} $ }
      \put(150,460){ $ ( q = u,d ) $ }
      \put(30,570){ $ {^{1} {\rm S}_{0}} $ }
      \put(52,570){ $ {^{3} {\rm S}_{1}} $ }
      \put(78,570){ $ {^{1} {\rm P}_{1}} $ }
      \put(100,570){ $ {^{3} {\rm P}_{J}} $ }
      \put(126,570){ $ {^{1} {\rm D}_{1}} $ }
      \put(148,570){ $ {^{3} {\rm D}_{J}} $ }
      \put(174,570){ $ {^{3} {\rm F}_{J}} $ }
      \put(51,455){ $ \pi $ }
      \put(72,485){ $ \rho $ }
      \put(0,437){$ {}_{3000} $}
      \put(0,392){$ {}_{2000} $}
      \put(0,347){$ {}_{1000} $}
      \put(13,302){$ {}_{0} $}
      \put(160,325){ $ s \bar{s} $ }
      \put(30,420){ $ {^{1} {\rm S}_{0}} $ }
      \put(52,420){ $ {^{3} {\rm S}_{1}} $ }
      \put(78,420){ $ {^{1} {\rm P}_{1}} $ }
      \put(100,420){ $ {^{3} {\rm P}_{J}} $ }
      \put(126,420){ $ {^{1} {\rm D}_{1}} $ }
      \put(148,420){ $ {^{3} {\rm D}_{J}} $ }
      \put(174,420){ $ {^{3} {\rm F}_{J}} $ }
      \put(51,329){ $ \eta_{s} $ }
      \put(72,344){ $ \phi $ }
      \put(216,437){$ {}_{3000} $}
      \put(216,392){$ {}_{2000} $}
      \put(216,347){$ {}_{1000} $}
      \put(229,302){$ {}_{0} $}
      \put(376,325){ $ q \bar{s} $ }
      \put(366,310){ $ ( q = u,d ) $ }
      \put(245,420){ $ {^{1} {\rm S}_{0}} $ }
      \put(267,420){ $ {^{3} {\rm S}_{1}} $ }
      \put(293,420){ $ {^{1} {\rm P}_{1}} $ }
      \put(315,420){ $ {^{3} {\rm P}_{J}} $ }
      \put(341,420){ $ {^{1} {\rm D}_{1}} $ }
      \put(363,420){ $ {^{3} {\rm D}_{J}} $ }
      \put(389,420){ $ {^{3} {\rm F}_{J}} $ }
      \put(266,319){ $ K $ }
      \put(287,337){ $ K^{\ast} $ }
      \put(-4,248){$ {}_{11000} $}
      \put(-4,200){$ {}_{10000} $}
      \put(0,152){$ {}_{9000} $}
      \put(160,165){ $ b \bar{b} $ }
      \put(44,270){ $ {^{1} {\rm S}_{0}} $ }
      \put(80,270){ $ {^{3} {\rm S}_{1}} $ }
      \put(125,270){ $ {^{1} {\rm P}_{1}} $ }
      \put(161,270){ $ {^{3} {\rm P}_{J}} $ }
      \put(188,245){ $ {}_{J} $ }
      \put(193,215){ $ {}_{2} $ }
      \put(188,212){ $ {}_{1} $ }
      \put(183,209){ $ {}_{0} $ }
      \put(193,197){ $ {}_{2} $ }
      \put(188,194){ $ {}_{1} $ }
      \put(183,191){ $ {}_{0} $ }
      \put(216,287){$ {}_{5000} $}
      \put(216,233){$ {}_{4000} $}
      \put(216,179){$ {}_{3000} $}
      \put(376,165){ $ c \bar{c} $ }
      \put(248,270){ $ {^{1} {\rm S}_{0}} $ }
      \put(273,270){ $ {^{3} {\rm S}_{1}} $ }
      \put(304,270){ $ {^{1} {\rm P}_{1}} $ }
      \put(329,270){ $ {^{3} {\rm P}_{J}} $ }
      \put(360,270){ $ {^{1} {\rm D}_{1}} $ }
      \put(385,270){ $ {^{3} {\rm D}_{J}} $ }
      \put(348,250){ $ {}_{J} $ }
      \put(348,213){ $ {}_{2} $ }
      \put(348,205){ $ {}_{1} $ }
      \put(348,197){ $ {}_{0} $ }
      \put(405,250){ $ {}_{J} $ }
      \put(405,241){ $ {}_{1} $ }
      \put(405,226){ $ {}_{2} $ }
      \put(405,218){ $ {}_{1} $ }
      \put(0,92){$ {}_{6000} $}
      \put(0,3){$ {}_{5000} $}
      \put(85,25){ $ q \bar{b} $ }
      \put(60,10){ $ ( q = u,d ) $ }
      \put(31,121){ $ {^{1} {\rm S}_{0}} $ }
      \put(53,121){ $ {^{3} {\rm S}_{1}} $ }
      \put(86,121){ $ {\rm P} $ }
      \put(180,15){ $ s \bar{b} $ }
      \put(123,121){ $ {^{1} {\rm S}_{0}} $ }
      \put(145,121){ $ {^{3} {\rm S}_{1}} $ }
      \put(178,121){ $ {\rm P} $ }
      \put(216,115){$ {}_{3000} $}
      \put(216,40){$ {}_{2000} $}
      \put(300,25){ $ q \bar{c} $ }
      \put(275,10){ $ ( q = u,d ) $ }
      \put(246,121){ $ {^{1} {\rm S}_{0}} $ }
      \put(268,121){ $ {^{3} {\rm S}_{1}} $ }
      \put(301,121){ $ {\rm P} $ }
      \put(395,15){ $ s \bar{c} $ }
      \put(338,121){ $ {^{1} {\rm S}_{0}} $ }
      \put(360,121){ $ {^{3} {\rm S}_{1}} $ }
      \put(393,121){ $ {\rm P} $ }
   \end{picture}
\caption{Quarkonium spectrum (lines represent experimental data).}
\label{figuussus}
\end{figure}

Circlets correspond to results of the type reported in
\cite{infrared}, but in which the hyperfine separation for the
1S and 2S
states has been evaluated up to the second order of perturbation
theory. In this case as running coupling constant we have taken
the Shirkov-Solovtsov expression (\ref{eq:runshk}). We have set again
$N_f=4$ and $\Lambda = 200$ MeV, but $\sigma =0.18\ {\rm GeV}^2$ and 
$m_s=0.39$ GeV, $ \; m_c=1.545$ GeV, $ \; m_b=4.898$ GeV. On the contrary
for the light quarks we have taken a phenomenological running mass in
terms of the modulus of the c.m. quark momentum, $m_u^2=m_d^2 = 
0.17 |{\bf k}| - 0.025|{\bf k}|^2 + 0.15 |{\bf k}|^4 \ {\rm  GeV}^2$ 
with |${\bf k}|$ in GeV. We can see that
in this way even the light pseudoscalar mesons turn out correctly, with
possibly some problems for the $q \bar b$ states and some other highly
excited states for which coupling with other channels
are probably important.


Squares correspond to preliminary
completely new calculations,
made using the analytic running constant (\ref{eq:runshk}),
running constituent masses $ m_{\rm P} (Q) $ (as described in Sec. 4)
for the light
and strange quarks, fixed masses
for the charm and the beauty quarks.
Inside $\alpha_{\rm s}(Q)$ the quantity $Q$ has been again
identified with
$|{\bf k}- {\bf k}^\prime|$. On the contrary,
for computational difficulties,
inside $m_{\rm P}(Q)$
we have taken
\begin{equation}
Q = {1 \over e^{\gamma_{\rm E}} \langle r \rangle} \,,
\end{equation}
\noindent
where $\gamma_{\rm E}$ is the Euler constant $ \gamma_{\rm E}= 
0.5772\dots $ and $\langle r \rangle $ is the radius 
of the unperturbed bound state \cite{schoberl}.
We have chosen
$N_f=3$, $\Lambda = 180$ MeV,
$\sigma =0.18 \, {\rm GeV}^2$,
$ { \sigma \over \pi } B_{\mu} = 0.48 $ GeV in
(\ref{eq:cnstmass}),
both for the light and the strange quarks,
and then
$\hat m_u = \hat m_d =25.0 $ MeV,
$\hat m_s = 87.3 $ MeV in (\ref{eq:runmass})
(in order to reproduce correctly the $\rho$ and the $\phi$ masses).
Finally we have taken $m_c = 1.508$ GeV and $m_b = 4.842$
for $c$ and $b$ quarks.
The results are not
of a better quality than
those obtained in the preceding calculation
but obviously conceptually more satisfactory.

As an example,
in the table
numerical values for the three types of calculations
are reported in the order
for the light-light channel.
For the third case
in the last column 
the pertinent values of running constituent light quark
mass are also reported for the various states.


\section{6. Conclusions}

In conclusion we can confirm what already noticed in references 
\cite{infrared} that our reduced  second order formalism together
with ansatz (\ref{eq:wilson}) can reproduce reasonably well the
general structure of the entire meson spectrum, light-light,
heavy-heavy and light-heavy sectors included. In order to obtain
the masses of light pseudo scalar mesons $\pi$, $ \eta_{s} $ and $K$,
however, a correct consideration of the infrared behavior
of the running coupling constant and of some kinds
of running constituent mass for the light quarks is
essential. The analytic Shirkov-Solovtsov coupling constant seems
to provide such a behavior.

What is new in this paper is the inclusion of second order
perturbative corrections to the hyperfine splitting in the case of
phenomenological running masses considered in \cite{infrared}
(circlets in Fig. 2) and the use of a running mass obtained
combining renormalization group and analyticity requirements with an 
approximate solution of the quark Dyson-Schwinger equation (squares in
Fig. 2).
\begin{theacknowledgments}
We are indebted to C. Simolo for
various calculations on the running masses.
\end{theacknowledgments}



\begin{thebibliography}{9}

\bibitem{quadratic}
  M. Baldicchi, G.M. Prosperi, {\sl Phys. Rev.} {\bf D 62}
  (2000) 114024;
  {\sl Fizika} {\bf B 8} (1999) 2, 251;
  {\sl Phys. Lett.} {\bf B 436} (1998) 145.
\bibitem{infrared}
  M. Baldicchi and G. M. Prosperi,  {\sl Phys. Rev.}
  {\bf D 66} (2002) 074008; {\sl Color confinement and hadrons
  Quantum Chromodynamics}, Page. 183, H. Suganuma, {\it et al.}
  eds. World Scientific 2004, hep-ph/0310213.
\bibitem{bmp}
  N. Brambilla, E. Montaldi, G.M. Prosperi, {\sl Phys. Rev.}
  {\bf D 54} (1996) 3506;
  G.M. Prosperi, 
  {\sl Problems of Quantum Theory of Fields}, Pag. 381,
  B.M. Barbashov, G.V. Efimov, A.V. Efremov Eds.
  JINR Dubna 1999,
  hep-ph/9906237.
\bibitem{simon}
  For a different approach to the relativistic bound state problem
  see : Yu.S. Kalashnikova, A.V. Nefediev, Yu.A. Simonov,
  {\sl Phys. Rev.} {\bf D 64} (2001) 014037;
  and references therein.
\bibitem{sanda}
  A.C. Mattingly, P.M. Stevenson, {\sl Phys. Rev.} {\bf D 49}
  (1994) 437;
  S. J. Brodsky, hep-ph/0412101;
  S. J. Brodsky, S. Menke, C. Merino,
  {\sl Phys. Rev.} {\bf D 67} (2003), 055008;
  P. Boucaud {\it et al.}, JHEP {\bf 04} (2000) 006;
  A. Ringwald and F. Schrempp {\sl Phys. Lett.} {\bf B 459}
  (1999) 24;
  Yu.L. Dokshitzer, A. Lucenti, G. Marchesini, G.P. Salam,
  {\sl JHEP} 9805 (1998) 003;
  Yu.L. Dokshitzer, V.A. Khoze, S.I. Troyan,
  {\sl Phys. Rev.} {\bf D 53} (1996) 89;
  N.V. Krasnikov, A.A. Pivovarov, {\sl Phys. Atom. Nucl.}
  {\bf 64} (2001) 1500;
  G. Grunberg, {\sl Phys. Rev.} {\bf D 46} (1992) 2228;
  A.I. Sanda {\sl Phys. Rev. Lett.}
  {\bf 42} (1979) 1658.
\bibitem{shirkov}
  D.V. Shirkov, I.L. Solovtsov, {\sl Phys. Rev. Lett.}
  {\bf 79} (1997) 1209;
  {\sl Theor. Math. Phys.} {\bf 120} (1999) 1220;
  D.V. Shirkov, hep-ph/0408272;
  A.I. Alekseev, B.A. Arbuzov, {\sl Mod. Phys. Lett.} {\bf A 13}
  (1998) 1747;
  A.I. Alekseev, {\sl Few Body Syst.} {\bf 32} (2003) 193;
  see also
  A. V. Nesterenko, {\sl Int. Journ. Mod. Phys.}
  {\bf A 18} (2003) 5475.
\bibitem{chiral}
  M.B. Hecht, C.D. Roberts, S.M. Schmidt,
  {\sl Phys. Rev.} {\bf C 63} (2001) 025213, and references therein.
\bibitem{barch}
 A. Barchielli, E. Montaldi, G.M. Prosperi,
  {\sl Nucl. Phys.} {\bf B 296} (1988) 625;
  Erratum-ibid. {\bf B 303} (1988) 752;
 A. Barchielli, N. Brambilla, G.M. Prosperi,
 {\sl Il Nuovo Cimento} {\bf 103 A} (1990) 59;
  N. Brambilla, P. Consoli, G.M. Prosperi,
  {\sl Phys. Rev.} {\bf D 50} (1994) 5878.
\bibitem{ginzburg}
  I.F. Ginzburg, D.V. Shirkov, {\sl Sov. Phys. JEPT} {\bf 22}
  (1966) 234.
\bibitem{data}
The Review of Particle Physics, S. Eidelman {\it et al.},
{\sl Phys. Lett.} {\bf B 592} (2004) 1.
\bibitem{schoberl}
 W. Lucha, F. Sch\"{o}berl, D. Gromes, {\sl Phys. Rep.}
 {\bf 200} (1991) 127.
\end{thebibliography}
\end{document}